\documentclass[11pt]{article}
\pdfoutput=1
\usepackage[utf8x]{inputenc}
\usepackage{amssymb,amsmath,mathrsfs,enumerate}
\usepackage{graphicx,rotate,multicol,braket, multirow}
\usepackage{cite}
\usepackage{braket}
\usepackage[normalem]{ulem}
\usepackage[colorlinks=true,
linkcolor=red,
urlcolor=blue,
citecolor=blue]{hyperref}
\usepackage{lineno}
\usepackage{color}
\DeclareMathOperator{\Tr}{Tr}
\def\ms#1{{\mathscr #1}}
\def\bar{\overline}
\def\tilde{\widetilde}
\def\uRdR{\begin{pmatrix} u'_R \\ d'_R \end{pmatrix}}

\evensidemargin=\oddsidemargin
\def\Eqn#1{Eq.\ (\ref{#1})}
\def\Eqs#1#2{Eqs.\ (\ref{#1}) and (\ref{#2})}
\def\Sect#1{Sec.\,\ref{#1}}
\title{\Large\bf 
Crossed two Higgs-doublet models: reduction of Yukawa parameters in the 
low-scale limit of left-right symmetry and other avatars}

\author{
  \sf Gustavo C.  Branco$^a$, 
  Dipankar Das$^b$,
  Miguel Levy$^a$,
  Palash B. Pal$^c$\thanks{e-mail:
    gbranco@tecnico.ulisboa.pt, 
    d.das@iiti.ac.in, miguelplevy@tecnico.ulisboa.pt,
    palashbaran.pal@saha.ac.in} \\[3mm]
  \small\em
  $^a$ Centro de F\'isica Te\'orica de Part\'iculas-CFTP and Departamento de
  F\'isica,  Instituto Superior T\'ecnico,\\  \small\em
  Universidade de Lisboa, Av
  Rovisco Pais, 1, P-1049-001 Lisboa, Portugal \\ 
    \small\em
    $^b$Indian Institute of Technology (Indore), Khandwa Road, Simrol,
  Indore 453 552, India \\ \small\em
  $^c$Department of Physics, University of Calcutta, 92 A. P. C. Road,
  Calcutta 700009, India}

\date{}

\begin{document}

%\pagewiselinenumbers

\maketitle

\renewcommand*{\thefootnote}{\arabic{footnote}}
\setcounter{footnote}{0} 
\begin{abstract}
We present new variants of the Two Higgs-Doublet Model where 
all Yukawa couplings with physical Higgs bosons are controlled by
the quark mixing matrices of both chiralities, as well as, in one case, the
ratio between the two scalar doublets' vacuum expectation values.  We
obtain these by imposing approximate symmetries on the Lagrangian
which, in one of the cases, clearly reveals the model to be the
electroweak remnant of the Minimal Left-Right Symmetric Model.  We also
argue for the benefits of the bidoublet notation in the Two Higgs-Doublet 
Model context for uncovering new models.
\end{abstract}

\bigskip
%===============================================

\section{Introduction} \label{s:intro}
%%%%%%%%%%
In the Standard Model (SM) of particle physics, the charged gauge
currents between quarks are guided by the Cabibbo-Kobayashi-Maskawa
(CKM) mixing matrix, while the neutral gauge currents are
flavor-diagonal.  The SM relies on the minimal choice of scalar fields
(one Higgs-doublet), and thus the quark mass matrices become
proportional to the corresponding Yukawa matrices.  Hence,
diagonalizing the quark mass matrices will automatically ensure the
simultaneous diagonalization of the Yukawa matrices.  Consequently, the
SM Higgs boson has only diagonal couplings, proportional to the quark
masses.

This straightforward picture may get perturbed even in the minimal
extensions beyond the SM~(BSM) such as the two Higgs-doublet
models~(2HDMs)\cite{Branco:2011iw, Bhattacharyya:2015nca}.  In a 2HDM,
the scalar sector of the SM is extended by adding a replica of the SM
Higgs-doublet.  As a result, there are two Yukawa matrices for fermions
of a given charge, and the diagonalization of the fermion mass
matrices will no longer guarantee the diagonalization of the Yukawa
matrices.  In other words, a 2HDM, in general, will contain flavor
changing neutral currents~(FCNCs) at the tree-level mediated by
neutral scalars.  Given that the FCNC couplings are, {\it a priori},
unknown, the analysis of the physical implications of a general 2HDM
may contain a lot of inherent arbitrariness.

As a simple way out, one tries to avoid the tree-level FCNCs
altogether by appropriate adjustments in the Yukawa sector.  The
simplest possibility was suggested by Glashow and Weinberg
\cite{Glashow:1976nt}.  According to their prescription, fermions 
of a specific charge should receive contributions to their mass from only
one of the scalar doublets.  In this way, similar to the SM, the Yukawa 
and the mass matrices for a particular species of fermions become 
proportional to each other thereby neutralizing the possibility of FCNC 
at the tree-level.  The general conditions for the absence of tree-level 
FCNCs in a 2HDM can be found in Refs.~\cite{Das:2018qjb, Botella:2018gzy}.

An interesting alternative to completely eliminating the tree-level
FCNCs is to accommodate them in a controlled manner.  This was achieved
by Branco, Grimus and Lavoura (BGL) \cite{Branco:1996bq}, where the
scalar FCNC couplings were related to the rows or columns of the CKM
matrix\cite{Bhattacharyya:2014nja, Botella:2014ska}.  In these BGL
models, flavored symmetries were introduced to appropriately texturize
the Yukawa matrices.  In this paper we make a similar effort to connect
the scalar FCNC couplings to the quark mixing parameters, thereby
reducing the arbitrariness in the Yukawa sector to a considerable
degree.  Yet, unlike the BGL models, we will rely on symmetries that
are completely flavor-blind, i.e., flavor universal.

Our current work also addresses the philosophical relevance of 2HDMs
in the present era.  A major part of the popularity of 2HDMs may be
attributed to minimal supersymmetry relying on a 2HDM scalar
structure.  However, current trends in the LHC Higgs data point towards
a not so bright future for minimal supersymmetry.  In such a case, one
may question the aesthetic appeal of 2HDM if it lacks the possibility
to be embedded in a more elegant theory.  However, a less known fact is
that the minimal left-right symmetric model~(LRSM) also results in a
2HDM Yukawa structure at the electroweak~(EW) scale \cite{Mohapatra:2019qid}.  
As we will show, such a 2HDM is very different from its canonical counterparts 
and can have quite distinct implications.

This article will be organized as follows.  In \Sect{s:genYuk} we
present a brief overview of the Yukawa sector in a general 2HDM,
following the usual conventions.  In \Sect{s:notation} an alternative
notation for the study 2HDMs is shown.  This is the usual notation of
the LRSM, which helps the connection between 2HDM and LRSM become
clear.  This notation is particularly helpful in uncovering new
models, which is done in \Sect{s:x2HDM}.  We include a phenomenological 
analysis in \Sect{s:Pheno}. Lastly, we summarize our
findings in \Sect{s:summary}.

%%%%%%%%%%%%%%%%%%%%%%%%%%%%%%%%%%%%%%%%%%%%%%%%%%%%%%%
%%%%%%%%%%  2HDM general Yukawa  %%%%%%%%%%%%%%%%%
%%%%%%%%%%%%%%%%%%%%%%%%%%%%%%%%%%%%%%%%%%%%%%%%%%%
\section{Yukawa sector in 2HDM: some generalities}
\label{s:genYuk}
%
%%%%%%%%%%%%%%%%%%%%
\subsection{Quark masses, mixings and couplings}
%%%%%%%%%%%%%%%%%%%%
We denote the quark fields in the original Lagrangian with primes:
\begin{eqnarray}
Q'_L = {u'_L \choose d'_L} \,, \qquad u'_R, \qquad d'_R \,,
%\label{}
\end{eqnarray}
where the generation index is suppressed.  The Higgs boson multiplets
$\phi_1$ and $\phi_2$ have a hypercharge assignment that yields the
following general Yukawa couplings:
\begin{eqnarray}
- \ms L_Y = \sum_{a=1}^2 \left[ \bar Q'_L \Gamma_a \phi_a d'_R + \bar
Q'_L \Delta_a \tilde \phi_a u'_R \right] + \mbox{h.c.,}
\label{e:LY}
\end{eqnarray}
where $\Gamma_a$ and $\Delta_a$ denote matrices in the generation
space, and
\begin{eqnarray}
\tilde \phi_a = i\sigma_2 \phi_a^* \,.
%\label{}
\end{eqnarray}
After spontaneous symmetry breaking~(SSB), we decompose the two
$SU(2)_L$ scalar doublets in their component form as follows: 
\begin{eqnarray}
\phi_a =\frac{1}{\sqrt{2}} \begin{pmatrix}
\sqrt{2} w_a^+ \\ v_a+h_a +iz_a
\end{pmatrix} \,, \qquad (a=1,2) \,.
\end{eqnarray}
We will assume that the vacuum expectation values~(VEVs) $v_1$ and
$v_2$ are real, and will use the notations
\begin{eqnarray}
v &=& \sqrt{v_1^2 + v_2^2} \,, \\
\tan\beta &=& v_2/v_1 \,.
%\label{}
\end{eqnarray}

After the spontaneous breaking of the gauge symmetry, the quarks
become massive.  The mass matrices are given by
\begin{subequations}
  \label{MdMu}
  \begin{eqnarray}
    M_d &=& \frac{1}{\sqrt{2}}\left( \Gamma_1 v_1 + \Gamma_2
    v_2\right) \,, \label{Md} \\*
    M_u &=& \frac{1}{\sqrt{2}} \left(\Delta_1 v_1 + \Delta_2
    v_2\right) \,. \label{Mu}
  \end{eqnarray}
\end{subequations}
These can be diagonalized through bi-unitary transformations:
\begin{subequations}
  \label{MtoD}
  \begin{eqnarray}
    U_{dL}^\dagger M_d U_{dR} &= D_d =& \mbox{diag} (m_d, m_s, m_b)
    \,,
    \label{MutoDu} \\
    U_{uL}^\dagger M_u U_{uR} &= D_u =& \mbox{diag} (m_u, m_c, m_t)
    \,.
    \label{MdtoDd}
  \end{eqnarray}
  \end{subequations}
As a result, by writing the charged current couplings of quarks in
terms of the physical fields, the combination
\begin{eqnarray}
V_L = U_{uL}^\dagger U_{dL} 
\label{VL}
\end{eqnarray}
emerges.  This is the CKM matrix.  Similarly, we can define a mixing
matrix for the right-handed quarks:
\begin{eqnarray}
V_R = U_{uR}^\dagger U_{dR} \,. 
%\label{}
\end{eqnarray}
Our aim in this paper is to search for models in which the Higgs
couplings to quarks are entirely determined by $V_L$ and $V_R$.

In order to discuss the Yukawa couplings, we first summarize the
spectrum of the scalar bosons.  The charged $(\omega^\pm)$ and the
neutral $(\zeta)$ Goldstone bosons can be extracted using the
following rotations
\begin{eqnarray}
	\begin{pmatrix} \omega^\pm \\ H^\pm	\end{pmatrix}
	= \begin{pmatrix} \cos\beta & \sin\beta \\ -\sin\beta & \cos\beta
	\end{pmatrix} \begin{pmatrix} w_1^\pm \\ w_2^\pm \end{pmatrix},
	\qquad
	\begin{pmatrix} \zeta \\ A	\end{pmatrix}
	= \begin{pmatrix} \cos\beta & \sin\beta \\ -\sin\beta & \cos\beta
	\end{pmatrix} \begin{pmatrix} z_1 \\ z_2 \end{pmatrix},
\end{eqnarray}
where, $H^\pm$ and $A$ stand for the physical charged scalar and
pseudoscalar respectively.  In the CP even sector, we apply the same
rotation to obtain
\begin{eqnarray}
\begin{pmatrix} H_0 \\ S	\end{pmatrix}
= \begin{pmatrix} \cos\beta & \sin\beta \\ -\sin\beta & \cos\beta
\end{pmatrix} \begin{pmatrix} h_1 \\ h_2 \end{pmatrix}.
\end{eqnarray}
The states $H_0$ and $S$ are not mass eigenstates in general.  However,
in the alignment limit\cite{Bhattacharyya:2013rya, Das:2015mwa,
  Dev:2014yca, Das:2019yad}, they become physical scalars and $H_0$
can be readily identified with Higgs scalar observed at the LHC
because it possesses SM-like couplings at the tree-level.  Thus the
quark couplings of $H_0$ are entirely flavor diagonal.  Without the
assumption of the alignment limit, the mass eigenstates would be
superpositions of $H_0$ and $S$, controlled by the parameters of the
scalar potential.  Hence, the quark couplings of the lightest scalar
field would not be flavor diagonal due to the $H_0$-$S$
mixture.  Nonetheless, assuming the alignment limit holds, only the
other neutral scalars, $S$ and $A$, can have flavor-changing couplings
to quarks, which will be an important theme in the subsequent
discussion.

It has been shown \cite{Botella:2009pq} that it is convenient to
define two matrices $N_d$ and $N_u$ as
\begin{subequations}
  \label{e:NdNu}
  \begin{eqnarray}
    N_d &=&\frac{1}{\sqrt{2}} U_{dL}^\dagger \Big( \sin\beta\,
    \Gamma_1 - \cos\beta\, \Gamma_2 \Big)
    U_{dR} \,, \label{e:Nd}\\
    N_u &=&\frac{1}{\sqrt{2}} U_{uL}^\dagger \Big( \sin\beta\,
    \Delta_1 - \cos\beta\, \Delta_2 \Big)
    U_{uR} \,, \label{e:Nu}
  \end{eqnarray}
whereby the quark couplings to the different Higgs bosons can be
written in the form
\begin{eqnarray}
  - \ms L_Y &=& \sqrt{2} \bigg[ \bar u \Big(
    N_u^\dagger V_L P_L -V_L N_d P_R \Big) dH^+ +
    \mbox{h.c.} \bigg] + \frac1v (\bar u D_u u + \bar d D_d d) H_0
  \nonumber\\*
  && -S \left\{\bar{d}\left(N_dP_R+N_d^\dagger P_L\right)d
  + \bar{u}\left(N_u P_R+N_u^\dagger P_L\right)u \right\} \nonumber \\
  && -iA \left\{\bar{d}\left(N_d P_R -N_d^\dagger P_L\right)d
  - \bar{u}\left(N_u P_R -N_u^\dagger P_L\right)u \right\}  \,,
\label{e:LYphys}
\end{eqnarray}
\end{subequations}
where $P_L$ and $P_R$ are the chirality projection operators.

%%%%%%%%%%%%%%%%%%%%%%%%%%%%%%%%%%%%%%%%%%%%%%%%%%%%%%%%%%%%%
%%%%%%%%%%%  General observation  %%%%%%%%%%%%%%%%%%%%%
%%%%%%%%%%%%%%%%%%%%%%%%%%%%%%%%%%%%%%%%%%%%%%%%%%%%%%%%%
\subsection{ Reducible Yukawa
  couplings }
%%%%%%%%%%
 From \Eqn{e:NdNu}, we see that the couplings of the Higgs bosons
 depend on the four diagonalizing matrices $U_{uL}$, $U_{dL}$,
 $U_{uR}$ and $U_{dR}$, as well as the matrices that appear in the
 Yukawa couplings.  We now show that there is a class of models in
 which the Yukawa couplings are {\em reducible}, by which we mean that
 the couplings are completely specified by the quark masses, and the
 left and right CKM matrices, $V_L$ and $V_R$.  The only dependence to
 the parameters of the Higgs potential is through the implicit
 dependence on the angle $\beta$.  Clearly, this requires $N_d$ and
 $N_u$ to be able to be written in terms of $V_L$ and $V_R$, apart
 from possible numerical factors.

The key to this reduction lies in the following observation.  Suppose,
in a given model, it is possible to write
\begin{subequations}
  \label{AB}
  \begin{eqnarray}
    \sin\beta\, \Gamma_1 - \cos\beta\, \Gamma_2 &=& \frac{\sqrt{2}}{v}
    (A_d M_d + B_d M_u) \,,
    \label{AdBd} \\
    \sin\beta\, \Delta_1 - \cos \beta\, \Delta_2 &=&
    \frac{\sqrt{2}}{v} (A_u M_u + B_u M_d) \,,
    \label{AuBu}
  \end{eqnarray}
\end{subequations}
with the numerical factors $A_d$, $B_d$, $A_u$, $B_u$.  Then
\Eqn{e:Nd} can be rewritten as
\begin{subequations}
  \label{NDreln}
  \begin{eqnarray}
    N_d &=& \frac1v U_{dL}^\dagger \Big( A_d M_d + B_d M_u \Big) U_{dR}
    \nonumber\\*
    &=& \frac1v \Big( A_d D_d + B_d V_L^\dagger D_u V_R \Big) \,.
    \label{NdD}
  \end{eqnarray}
  Similarly, from \Eqn{e:Nu} one obtains
  \begin{eqnarray}
    N_u &=& \frac1v \Big( A_u D_u + B_u V_L D_d V_R^\dagger \Big) \,.
    \label{NuD}
  \end{eqnarray}
\end{subequations}
Therefore, Yukawa couplings will be completely determined by the quark
masses and mixing matrices if \Eqn{AB} holds.

However, it should be clear that it is not possible to write relations
of the form of \Eqn{AB} in the most general case.  Four independent
matrices, $\Gamma_a$ and $\Delta_a$ for $a=1,2$, cannot be written in
terms of two matrices $M_d$ and $M_u$.  Therefore, it is necessary to
have only two independent Yukawa matrices.  In order to achieve this
goal, it is necessary to introduce some condition to restrict the
Yukawa matrices.  Later in this paper, we discuss some such relations,
and the resulting Yukawa couplings.

We noticed in \Eqn{e:LYphys} that the couplings of the neutral Higgs
bosons, $S$ and $A$, to the up-type and down-type quarks are governed
by the matrices $N_u$ and $N_d$ respectively.  From \Eqn{NDreln}, we
see that the parts $A_u$ and $A_d$ are proportional to the diagonal
mass matrices in the respective sector, and are therefore flavor
diagonal.  Thus, FCNC occurs only through the parts $B_u$ and $B_d$,
and are absent in a model where these parts vanish.  In such models,
the Higgs couplings are even independent of the quark mixing matrices.
The conventional type-I and type-II 2HDMs constitute examples of this
category, which will be discussed in \Sect{s:oldtypes}.  But the aim
of this paper is to uncover other interesting models where \Eqn{AB}
holds, and the Yukawa couplings are governed by \Eqn{NDreln}.

%%%%%%%%%%%%%%%%%%%%%%%%%%%%%%%%%%%%%%%%%%%%%%%%%%%%%%%
%%%%%%%%%%  Bidoublet notation  %%%%%%%%%%%%%%%%%%
%%%%%%%%%%%%%%%%%%%%%%%%%%%%%%%%%%%%%%%%%%%%%%%%%%%
\section{A notational digression}
\label{s:notation}
In order to find nontrivial examples of 2HDMs where \Eqn{AB} holds we
find it convenient to write the two doublets together, into what can
be called a bidoublet:
\begin{eqnarray}
\Phi = \begin{pmatrix} \tilde{\phi}_1 & \phi_2 \end{pmatrix}  \,.
\label{Phi}
\end{eqnarray}
The transformation properties of the Higgs-doublets under the SM gauge
symmetry can be expressed in a concise manner using the bidoublet:
\begin{eqnarray}
\Phi \xrightarrow{\ {\rm SU(2)}_L\times {\rm U(1)}_Y\ } \Sigma_L\, \Phi\,
e^{-\frac{i}{2}\sigma_3 \theta(x)} \,, 
\label{transf}
\end{eqnarray}
where $\Sigma_L$ denotes an element of ${\rm SU(2)}_L$ and the
appearance of $\sigma_3$ on the right takes care of the fact that the
hypercharges of $\phi_k$ and $\tilde{\phi}_k$ are opposite.  It should
now be noted that one can construct additional bidoublets as well, all
of which have the same transformation properties under ${\rm
  SU(2)}_L\times {\rm U(1)}_Y$ as $\Phi$:
\begin{subequations}
  \label{likePhi}
  \begin{eqnarray}
    \tilde{\Phi} =& \sigma_2 \Phi^* \sigma_2 &\equiv \begin{pmatrix}
      \tilde{\phi}_2 & \phi_1	\end{pmatrix} \,, \\
    \Psi =& \Phi\, \sigma_3 &\equiv \begin{pmatrix}
      \tilde{\phi}_1 & -\phi_2	\end{pmatrix} \,, \\
    \tilde\Psi =& \sigma_2 \Psi^* \sigma_2
    &\equiv \begin{pmatrix} -\tilde{\phi}_2 & \phi_1	\end{pmatrix}
    \,.
  \end{eqnarray}
\end{subequations}

In keeping with the bidoublet notation for the Higgs multiplets, the
right-handed quark fields can be written in a column with two
components.  Note that the gauge transformation on this column can
also be written in a succinct form:
\begin{eqnarray}
\uRdR \xrightarrow{\ {\rm SU(2)}_L\times {\rm U(1)}_Y\ }
e^{+\frac{i}{6}\theta(x)} e^{+\frac{i}{2}\sigma_3 \theta(x)} \uRdR \,,
\label{uRdR}
\end{eqnarray}
whereas the transformation of the left-handed quark doublets are given by
\begin{eqnarray}
Q'_L  \xrightarrow{\ {\rm SU(2)}_L\times {\rm U(1)}_Y\ }
\Sigma_L \, e^{+\frac{i}{6}\theta(x)}\, Q'_L \,.
%\label{}
\end{eqnarray}

The four different Yukawa coupling matrices that appeared in
\Eqn{e:LY} are now encrypted in the couplings of the quarks with these
four different bidoublets given in \Eqs{Phi}{likePhi}:
\begin{eqnarray}
- \ms L_Y = \left[Y_\Phi \bar{Q}_L \Phi \uRdR +
\tilde Y_\Phi \bar{Q}_L \tilde{\Phi} \uRdR + 
Y_\Psi \bar{Q}_L \Psi \uRdR + 
\tilde Y_\Psi \bar{Q}_L \tilde\Psi \uRdR
\right] + \mbox{h.c.} \,.
\label{e:LYbi}
\end{eqnarray}
Comparing \Eqs{e:LY}{e:LYbi}, it is easy to see the relations between
the two different sets of notations:
\begin{subequations}
  \label{e:connection}
  \begin{align}
    \Gamma_1 &= \tilde Y_\Phi + \tilde Y_\Psi \,, & \Gamma_2 &= Y_\Phi
    - Y_\Psi \,, 	\\
    \Delta_1 &= Y_\Phi + Y_\Psi \,, & \Delta_2 &= \tilde Y_\Phi -
    \tilde Y_\Psi \,.
    %\label{}
  \end{align}
\end{subequations}
%%

%%%%%%%%%%%%%%%%%%%%%%%%%%%%%%%%%%%%%%%%%%%%%%%%%%%%%%%%%
%%%%%%%%%%  The Model   %%%%%%%%%%%%%%%%%%%%%
%%%%%%%%%%%%%%%%%%%%%%%%%%%%%%%%%%%%%%%%%%%%%%%%%%%%%%%%
%
\section{The crossed 2HDMs} 
\label{s:x2HDM} 
We will now proceed to construct nontrivial examples of 2HDMs where
\Eqn{AB} holds.  But first, let us recover the conventional 2HDMs
which prevent any FCNC at the tree level.

\subsection{Retrieving the type-I and type-II 2HDMs}\label{s:oldtypes}
In type-I 2HDM, only $\phi_1$ is odd under a $Z_2$ symmetry while all
other fields are even.  Consequently, only $\phi_2$ couples to all the
fermions.  In the bidoublet notation, we can write this $Z_2$ symmetry
as 
\begin{eqnarray}
	\Phi \to - \Phi \sigma_3 \,.
	\label{e:sym1}
\end{eqnarray}
The above transformation will affect the remaining bidoublet
structures as
\begin{eqnarray}
 \Psi \to -\Psi \sigma_3, \qquad  \tilde{\Phi} \to \tilde{\Phi}
 \sigma_3, \qquad  \tilde{\Psi} \to \tilde{\Psi} \sigma_3.  
\end{eqnarray}
The Yukawa Lagrangian of \Eqn{e:LYbi} will remain unaffected by the above
transformation if
\begin{eqnarray}
  Y_\Phi = -Y_\Psi \quad {\rm and} \quad \tilde Y_\Phi
  = -\tilde Y_\Psi \,,
\end{eqnarray}
which, in view of \Eqn{e:connection}, implies
\begin{eqnarray}
    \Gamma_1 = \Delta_1 = 0 \,.
\end{eqnarray}
It is easy to see that in this model, $A_u=A_d=-\cot\beta$,
$B_u=B_d=0$.  Since the $B$ coefficients are zero, there is no FCNC in
this model.

In type-II 2HDM, $\phi_1\to -\phi_1$ and $d'_R\to -d'_R$ under the
$Z_2$ symmetry.  Thus, $\phi_1$ will couple only to the down-type quarks
whereas $\phi_2$ will couple to the up-type quarks.  This can be
ensured via the following transformations in the bidoublet notation:
\begin{eqnarray}
 \Phi \to - \Phi \sigma_3 & \mbox{and} & \uRdR \to \sigma_3 \uRdR \,.
\label{e:sym2}
\end{eqnarray}
It is then easily seen that to keep the Yukawa Lagrangian of
\Eqn{e:LYbi} invariant under the above transformations, we must
require
\begin{eqnarray}
Y_\Phi = Y_\Psi = 0 \,,
\end{eqnarray}
which, in view of \Eqn{e:connection}, translates into
\begin{eqnarray}
\Delta_1 = \Gamma_2 = 0 \,.
\end{eqnarray}
This means that $A_d=\tan\beta$, $A_u=-\cot\beta$, $B_u=B_d=0$ in this
model.

Note that we could have defined the $Z_2$ symmetry differently, by
omitting the minus sign in the transformation law of right-handed
quarks from \Eqn{e:sym2}.  That would not have given us a new model:
it would have just interchanged the roles of $\phi_1$ and $\phi_2$.

None of the examples presented above belong to the class that we call
``crossed 2HDM'' or ``x2HDM'', for reasons that we will explain
shortly.  They come next.

%%%%%%%%%%
\subsection{ First example of crossed 2HDM~: connection with
  left-right symmetry}\label{s:x2hdm1}
%%%%%%%%%%
Consider a symmetry under which the nontrivial transformations are
\begin{eqnarray}
\Phi \to \Phi \Sigma_R^\dagger \,, \qquad \uRdR \to \Sigma_R \uRdR \,,
\label{symm2a}
\end{eqnarray}
where $\Sigma_R$ is any SU(2) matrix.  Since $2\times2$ unitary
matrices have the property
\begin{eqnarray}
\Sigma_R^* = - \sigma_2\, \Sigma_R \, \sigma_2 \,, 
%\label{}
\end{eqnarray}
it is easily seen that, under the transformation of \Eqn{symm2a},
$\tilde\Phi$ transforms the same way as $\Phi$, but $\Psi$ and
$\tilde\Psi$ do not, because of the presence of a factor of $\sigma_3$
in their definitions.  Thus, the Yukawa couplings associated with
$\Psi$ and $\tilde\Psi$ are not invariant under this symmetry.  It
should be noted that this symmetry should be considered as an
approximate symmetry, since it does not commute with the hypercharge
symmetry.  Imposing the symmetry of \Eqn{symm2a} on the Yukawa
Lagrangian of \Eqn{e:LYbi}, we will obtain the following restrictions
on the Yukawa matrices:
\begin{eqnarray}
Y_\Psi = \tilde Y_\Psi = 0 \,,
\label{cond2aY}
\end{eqnarray}
leading to
\begin{eqnarray}
\Gamma_1 = \Delta_2 \equiv \Gamma ~~({\rm say}) \,, \qquad \Gamma_2 =
\Delta_1 \equiv \Delta ~~({\rm say}) \,.
\label{e:x2hdm-1}
\end{eqnarray}
In this case we will have the following mass matrices
\begin{eqnarray}
M_d = \frac{v}{\sqrt{2}}\left(\cos\beta\, \Gamma + \sin\beta\,
\Delta\right) \,, \qquad 
M_u = \frac{v}{\sqrt{2}}\left(\cos\beta\, \Delta + \sin\beta\,
\Gamma\right) \,.  
\label{e:mx1}
\end{eqnarray}
Inverting these equations and comparing with \Eqn{AB}, one obtains
\begin{eqnarray}
A_d = A_u = \tan 2\beta \,, \qquad B_d = B_u = - \sec2\beta\,.  
%\label{}
\end{eqnarray}
Plugging this into the definitions \Eqn{NDreln}, one obtains
\begin{subequations}
  \label{e:NuNdx1}
  \begin{eqnarray}
    N_d = \frac1v \left( \tan2\beta \; D_d - \sec2\beta \; V_L^\dagger
    D_u V_R \right) \,, \\
    N_u = \frac1v \left( \tan2\beta \; D_u - \sec2\beta \; V_L
    D_d V_R^\dagger \right) \,.
  \end{eqnarray}
\end{subequations}
As such, the FCNC couplings of the neutral Higgs bosons are fully
controlled by the quark mixing parameters and $\tan\beta$.  This is a
crossed 2HDM, which we will call x2HDM-1 in subsequent discussion.

The symmetry of \Eqn{symm2a}, which was used to arrive at this model,
is qualitatively different from those introduced in \Sect{s:oldtypes}.
The point is that the transformations produce linear superpositions of
the SM doublets $\tilde\phi_1$ and $\phi_2$.  Since these two objects
have opposite hypercharges, such mixing is not allowed by gauge
symmetry.  So, a symmetry of this sort can be imposed on the Yukawa 
sector only, although it will be violated by the gauge interactions, and 
therefore can only be an approximate symmetry of the full Lagrangian.  
We call these {\em crossed} symmetries because it connects across 
different hypercharges.

However, the particular transformations of \Eqn{symm2a} can easily be
promoted to be a symmetry of the full Lagrangian.  These
transformations are easily seen as the transformations of the relevant
fields under an ${\rm SU(2)}_R$ symmetry.  Thus, in effect, the imposition
of the symmetry of \Eqn{symm2a} implies that the Yukawa couplings have
a symmetry ${\rm SU(2)}_L \times {\rm SU(2)}_R \times U(1)$, which is
the gauge symmetry of the LRSMs \cite{Mohapatra:1974gc, Mohapatra:1974hk,
 Senjanovic:1975rk}.  We can therefore extend the symmetry to the entire 
Lagrangian and build a LRSM.  In fact, our Yukawa couplings are no different than 
the usual ones encountered in the LRSMs that involve a bidoublet 
Higgs multiplet $\Phi$ transforming as the $(2,2,0)$ representation of
the gauge group.  In the context of LRSMs, it was noted
\cite{Deshpande:1990ip} that the fermion couplings with Higgs bosons
depend only on $V_L$ and $V_R$.

%%%%%%%%%%%%%%%%%%%%%%%%%%%%%%%%%%%%%%%%%%%%%%%%%%%%%%%%%%%%%%%%
%%%%%%%%%%%%%%  x2HDM-2  %%%%%%%%%%%%%%%%%%%%%%%%%
%%%%%%%%%%%%%%%%%%%%%%%%%%%%%%%%%%%%%%%%%%%%%%%%%%%%%%%%%
\subsection{ More examples of crossed 2HDM}
So far, our approach may appear as a convoluted exercise to connect
the LRSM with 2HDM.  However, the notations that we adopted in this
paper can be used to uncover new types of 2HDMs which were previously
unknown.

As an example, we introduce a $Z_2$ symmetry in the following
form:
\begin{eqnarray}
\Phi \to \Phi\, \sigma_1 \,, \qquad \uRdR \to \sigma_1 \uRdR \,.
\label{e:symx2}
\end{eqnarray}
Note that, this also does not commute with the hypercharge symmetry,
and therefore should be considered as an approximate symmetry.
This symmetry, when imposed on the Yukawa Lagrangian of \Eqn{e:LYbi},
implies the following
\begin{eqnarray}
\tilde Y_\Phi = Y_\Psi = 0 \,,
\label{cond2bY}
\end{eqnarray}
which means 
\begin{eqnarray}
\Gamma_1 = - \Delta_2 \equiv \Gamma \quad ({\rm say}) \,, \qquad \Gamma_2
= \Delta_1 \equiv \Delta \quad ({\rm say}) \,.
\label{e:Yx2}
\end{eqnarray}
This model will be called x2HDM-2.  As a consequence of \Eqn{e:Yx2},
the quark mass matrices will now become,
\begin{eqnarray}
M_d = (\cos\beta\, \Gamma + \sin\beta\, \Delta)v/\surd2 \,, \qquad
M_u = (\cos\beta\, \Delta - \sin\beta\, \Gamma)v/\surd2 \,.
\label{e:mx2}
\end{eqnarray}
Inverting these equations and comparing with \Eqn{AB}, one obtains
\begin{eqnarray}
A_d = A_u = 0 \,, \qquad B_u = - B_d = 1 \,.
%\label{}
\end{eqnarray}
As a result, the matrices $N_u$ and $N_d$ are given by
\begin{subequations}
  \label{e:NuNdx2}
\begin{eqnarray}
N_d &=& - \frac1v V_L^\dagger D_u V_R \,, \\
N_u &=& \frac1v V_L D_d V_R^\dagger \,.
\end{eqnarray}
\end{subequations}
This is an intriguing case where the Yukawa couplings with physical
Higgs bosons are independent of $\tan\beta$, the ratio of the two VEVs.

One may consider other relations among the Yukawa matrices $Y_\Phi$,
$\tilde Y_\Phi$, $Y_\Psi$ and $\tilde Y_\Psi$, which can potentially
give rise to different structures of $N_u$ and $N_d$.  Not all
relations will produce new models.  For example, changing $\sigma_1$
to $\sigma_2$ in \Eqn{e:symx2} produces the same restrictions on
Yukawa couplings as those shown in \Eqn{e:Yx2}.  Some other conditions
might result in equations which imply only an interchange of the names
$\phi_1$ and $\phi_2$, and therefore a redefinition of $\beta$.  But
there is no reason why more models cannot be produced which have
different physical implications.  However, it is not always
straightforward to motivate arbitrary relations between the Yukawa
matrices from symmetries.

%%%%%%%%%%%%%%%%%%%%%%%%%%%%%%%%%%%%%%%%%%%%%%%%%%%%%%%%%
%%%%%%%%%%%  A digression on symmetries  %%%%%%%%%%%%%%
%%%%%%%%%%%%%%%%%%%%%%%%%%%%%%%%%%%%%%%%%%%%%%%%%%%%%%%
\subsection{Some specificities on the x2HDMs}
It has been pointed out that, unlike the symmetries in
\Eqs{e:sym1}{e:sym2}, the ones shown in \Eqs{symm2a}{e:symx2} mix
fields with different hypercharges.  Therefore, these symmetries do
not commute with the $U(1)_Y$ part of the SM gauge symmetry.  Thus, as
previously stated, it should be considered as an approximate symmetry,
imposed only on the Yukawa sector, and can prevail in the Lagrangian
only in the limit when the $U(1)_Y$ gauge coupling ($g'$)
vanishes.  This approximate character or, in other words, the
interference with the SM hypercharge gauge group can be explicitly
seen through the computation of the renormalization group equations
of the Yukawa couplings.  If the relations of \Eqn{e:x2hdm-1}
or \Eqn{e:Yx2} are imposed at a certain scale, then they will evolve 
with the change of scale according to the formulas \cite{Branco:2011iw}
                                %\footnote{Here, we
                                %extend the SM through the inclusion
                                %of right-handed neutrinos, and impose
                                %the relations of \Eqn{e:x2hdm-1} on
%the leptonic sector as well.}
%
\begin{eqnarray}
	16 \pi^2 \frac{d}{d\ln \mu} (\Delta_1-\Gamma_2) =-g^{\prime
          2}\Delta \,, \qquad  16 \pi^2 \frac{d}{d\ln \mu}
        (\Gamma_1\mp\Delta_2)= g^{\prime 2}\Gamma \,,  
\end{eqnarray}
where we assume the presence of right-handed neutrinos with appropriate 
Yukawa interactions involving the doublet Higgs bosons, and extend the 
symmetry to the leptonic sector.

\begin{figure}
\centering
\begin{minipage}{.5\textwidth}
  \centering
  \includegraphics[width=.9\linewidth]{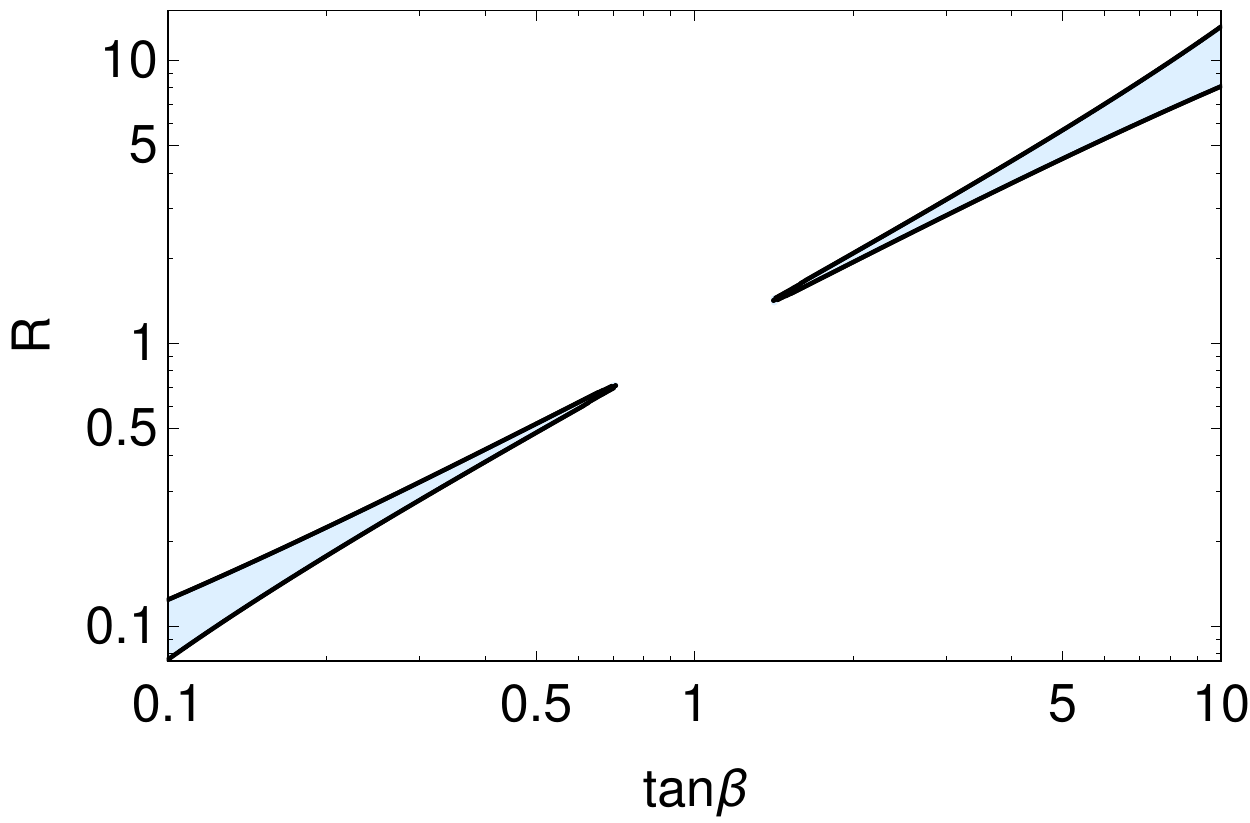}
\end{minipage}%
\begin{minipage}{.5\textwidth}
  \centering
  \includegraphics[width=.9\linewidth]{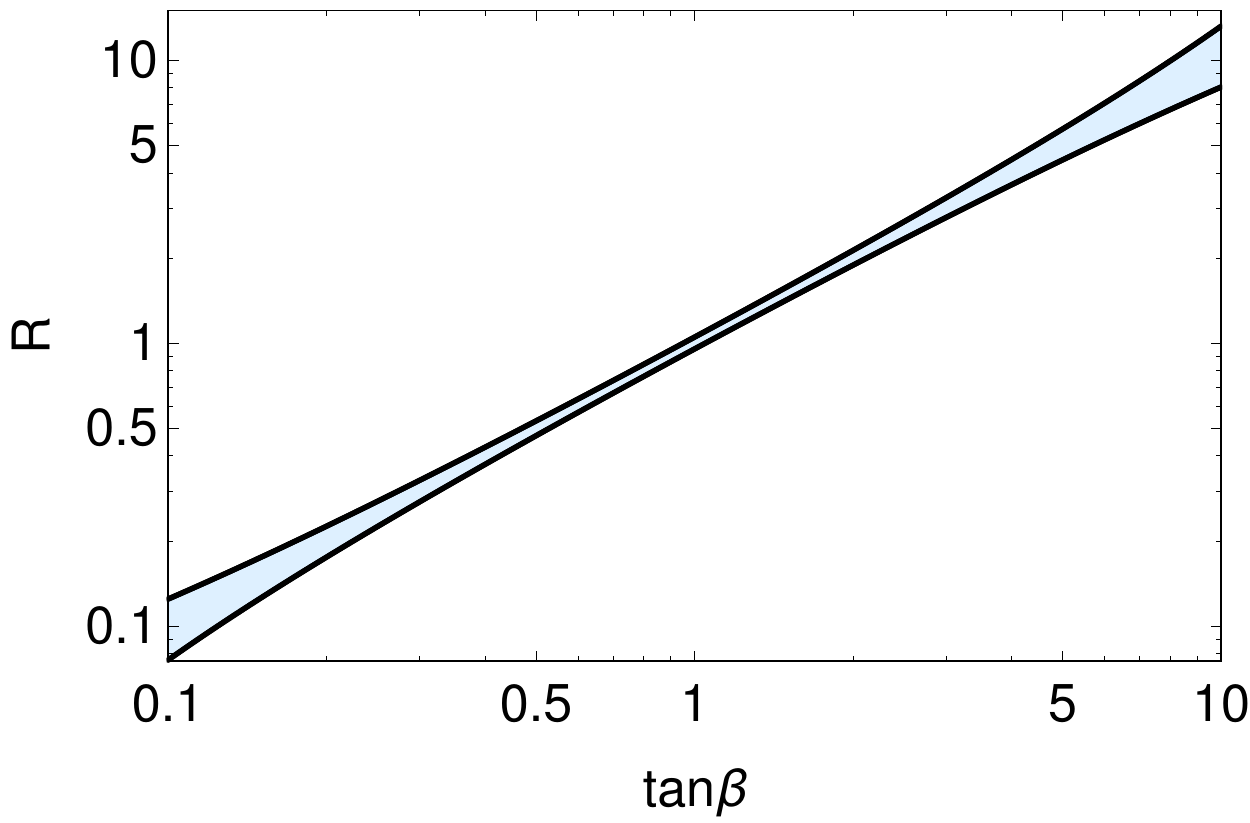}
\end{minipage}
\caption{\small Plot of $\tan\beta$ vs
  $R=\sqrt{\Tr(\Gamma^\dagger\Gamma)/\Tr(\Delta^\dagger\Delta)}$ for
  randomly generated $\Gamma$ and $\Delta$. The shaded region is consistent 
  with the observed quark masses and mixings, in the x2HDM-1 (left) and x2HDM-2
  (right).  We impose a perturbativity limit of $|\Gamma_{ab}|,
  |\Delta_{ab}|\le 1$.}
\label{f:corr}
\end{figure}

%%%%%%%%%%%%%%%%%%%%%%%%%%%%%%%%%%%%%%%%%%%%%%%%%%%%%%%%%
%%%%%%%%%%%%  A Direct Consequence   %%%%%%%%%%%%%%%%%%
%%%%%%%%%%%%%%%%%%%%%%%%%%%%%%%%%%%%%%%%%%%%%%%%%%%%%%%%
%\subsubsection{A direct consequence}
%
By taking a closer look at the relations of both x2HDMs, it is
possible to extract one characteristic which is general to all
x2HDMs.  Suppose the inversion of \Eqn{MdMu} yields the solutions
\begin{eqnarray}
  \Gamma = p_1 M_u + p_2 M_d \,, \qquad \Delta = q_1 M_u + q_2 M_d \,,
  \label{pq}
\end{eqnarray}
for some assignment of $\Gamma$ and $\Delta$ from among the four
Yukawa matrices.  We can now form the traces of the hermitian matrices
$\Gamma^\dagger \Gamma$ and $\Delta^\dagger \Delta$, each of which will
contain four terms.  Since $\Tr(M_u^\dagger M_u) \approx m_t^2$, we
expect this term to dominate.  If it indeed does, then
\begin{eqnarray}
    {\Tr(\Gamma^\dagger \Gamma) \over \Tr (\Delta^\dagger \Delta)} =
    {p_1^2 \over q_1^2} + \mbox{(small terms.)}
    %\label{}
\end{eqnarray}
This means that there should a strong correlation between the square
root of the left side of this equation and $|p_1/q_1|$.  For all
x2HDMs presented here, $|p_1/q_1| = |\tan\beta|$.  The correlation is
shown in Fig.\,\ref{f:corr}.

What we do to plot the graphs is this.  We take the quark masses as
given, and also the components of $V_L$.  We then randomly generate
$U_{uL}$, $U_{uR}$ and $U_{uR}$, find $U_{dL}$ from \Eqn{VL}, and,
with each random choice, generate $M_d$ and $M_u$ from \Eqn{MtoD} and
use \Eqn{pq}, as applicable to a particular model, to find $\Gamma$
and $\Delta$.  These are then used to make the plot, and the
correlation clearly shows.

We notice from Fig.~\ref{f:corr} the weakening of the correlation as
we move away from $\tan\beta=1$.  This can be understood as a direct
consequence of the strong hierarchy between the up and down-quark
masses.  For $\tan\beta \approx 1$, we need to arrange a cancellation
in the expression for $M_d$ to reproduce such a strong hierarchy.  This
will approximately fix $\tan\beta$.  However, for $\tan\beta$ far away
from unity ({\it i.e.}, for either $\sin\beta$ or $\cos\beta$ close to
zero), the matrices $\Gamma$ and $\Delta$ in \Eqs{e:mx1}{e:mx2}
effectively serve as independent sources of masses for the up and
down-type quarks.

One particular aspect of the x2HDM-1 can easily be seen in by looking
at \Eqn{e:mx1}.  Namely, for $\tan\beta=1$ we will have $M_u=M_d$
leading to unacceptable phenomenological results.  Therefore we must be
away from $\tan\beta=1$ to reproduce realistic values for the physical
quark masses and mixings.  Additionally, problems in the region
surrounding $\tan\beta=1$ can be understood by inverting \Eqn{e:mx1}
to obtain $\Gamma$ and $\Delta$ in terms of $M_u$ and $M_d$.  These
expression will have terms proportional to $\sec2\beta$, which is
large near the $\tan\beta=1$ region, leading to non-perturbative
Yukawa couplings.  One may then naturally wonder how close can $\tan\beta$ be to
unity so that the observed values of the quark masses and mixings are
recovered while, at the same time, the elements of Yukawa matrices in
\Eqn{e:x2hdm-1} are kept under the perturbative limit, $|\Gamma_{ab}|,
|\Delta_{ab}|\le 1$.  From Fig.~\ref{f:corr} (left), we can read the
forbidden region in $\tan\beta$ as follows:
\begin{eqnarray}
	0.75 \lesssim \tan\beta \lesssim 1.33 \,.
\end{eqnarray}
We argued earlier that the model presented in \Sect{s:x2hdm1} is the
low-energy limit of the left-right symmetric model.  In this
connection, it should be pointed out that our results on $\tan\beta$
are equally applicable in the case of LRSM where $\tan\beta$ will
obviously be redefined as the ratio of the two VEVs of the
bidoublet.  It should also be noted that despite the LRSM being in
existence for decades, such constraints on $\tan\beta$ have not been
emphasized earlier.

\section{Phenomenology of the x2HDMs}
\label{s:Pheno}

Our goal was to relate the FCNC parameters to the LH and RH quark mixing parameters. Having achieved that goal, we now briefly turn our attention to the consequences of experimental constraints on the models. More specifically, by relating $N_d$ with $V_R$, we greatly reduce the free parameters of the model, yet these FCNC contributions are still present at the tree-level. As such, as our first objective, we set out to neutralize these contributions to minimize the impact they have on neutral meson mixing. We expect this to lead to a very constrained $V_R$, due to the high experimental precision of $\Delta M_{P}$, where $P=K, B_s, B_d$. Interestingly, the coupling structure of the x2HDMs is such that these are the same couplings that drive the fermionic decays of the nonstandard scalars of the model. Thus, by finding one $V_R$ compatible with $\Delta F=2$ flavor observables, the models will have a distinct prediction for the ratio of fermionic non-SM scalar decays: $\text{Br}\left( S, A \rightarrow \bar f f\right)/\text{Br}\left( S, A \rightarrow \bar b b\right)$. As mentioned earlier, we work under the assumption of the alignment limit, where $H_0$ is a SM-like Higgs particle with flavor-diagonal couplings. Thus, all NP FCNC come exclusively from $S$ and $A$.

To tame the tree-level effects of the nonstandard scalars in $\Delta M_P$, we first write the relevant expression for the NP contribution to the meson mass difference ($\Delta M^{\rm NP}_P$) as \cite{Atwood:1996vj, Nebot:2015wsa}
\begin{eqnarray}
2 M_P \Delta M_P^{\rm NP} &=& \left| \left(\frac{1}{M_A^2}-\frac{1}{M_S^2}\right) \left[ {\left({\left(N_d^*\right)}_{ji}\right)}^2 + {\left({\left(N_d\right)}_{ij}\right)}^2\right] \frac{5}{3}M_P^{0,F}  \right. \nonumber \\ 
&&\left.- \left(\frac{1}{M_A^2}+\frac{1}{M_S^2}\right) 2 {\left(N_d\right)}_{ij}{\left(N^*_d\right)}_{ij} \left(\frac{M_A^{0,F}}{3} -2M_P^{0,F}\right) \right|,
\label{eq:DeltaM}
\end{eqnarray}
where $P= \bar q_i q_j$ , and 
\begin{equation}
M_P^{0,F} = -f_P^2 \frac{M_P^4}{\left(m_{q_i}+m_{q_j}\right)^2}, \qquad M_A^{0,F} = f_P^2 M_P^2.
\end{equation}
In the above, $m_{q_i}$ is the mass of the quark $q_i$, whereas $f_P$ and $M_P$ are the decay constant and the mass of the meson $P$, respectively.

Clearly, in the limit $M_S=M_A$, there is a cancellation in the first term of \Eqn{eq:DeltaM}. In order to sufficiently dilute the contribution of the second term of \Eqn{eq:DeltaM}, we require, as an example, ${\left(N_d\right)}_{21}\sim{\left(N_d\right)}_{31}\sim{\left(N_d\right)}_{23}\sim 0$, which leads to $\Delta M^{\rm NP}_P \sim 0$. Ignoring, for now, the possible phases of $V_R$, we can constrain the three Euler angles through the three conditions above. This fixes $V_R$ to a very precise degree, where one specific example is (assuming the Wolfenstein parametrization of $V_L$ \cite{Wolfenstein:1983yz}):
\begin{equation}
V_R\approx 
\begin{pmatrix}
 1 & 6.65 \times 10^{-5}& 3.86 \times 10^{-4} \\
 -3.92\times 10^{-4} &0.169 & 0.986 \\
 1.20 \times 10^{-7} & -0.986& 0.169 
\end{pmatrix}.
\label{RHckm}
\end{equation}
Using this $V_R$, we have explicitly checked that the $\Delta F=2$ contributions to $K$, $B_s$, and $B_d$ oscillations are under control for $M_S=M_A\sim \mathcal{O}(\text{TeV})$. It is interesting to note that, in this particular example, some of the off-diagonal terms are quite large.

\begin{figure}[h]
\centering
\includegraphics[scale=.5]{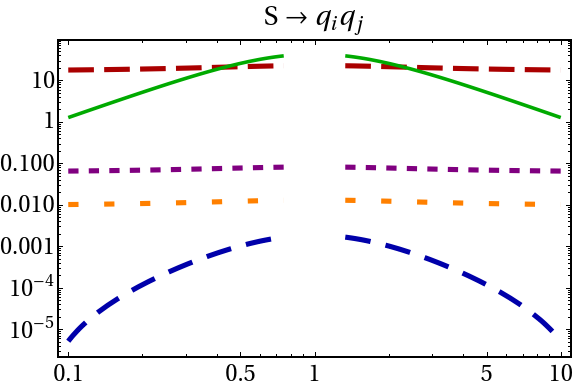}%~~~\includegraphics[scale=.4]{decaysA} 
\\
\includegraphics[width=.6\linewidth]{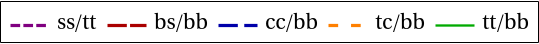}
\caption{\small Log-log plot of relative branching ratios for the $S$ decay into quark-antiquark pairs, normalized by its branching ratio into a $\bar bb$  quark pair, as a function of $\tan\beta$, for the x2HDM-1, for $M_S=M_A=1.5 \text{ TeV}$. The region for $\tan\beta$ which was excluded in Figure~\ref{f:corr} is intentionally kept out in these plots. The relative branching ratios for $A$ are very similar.}
\label{fig:decays}
\end{figure}

Now that we have established that TeV-scale nonstandard scalars can successfully negotiate the stringent $\Delta F=2$ flavor constraints, it is interesting to find distinctive features of these scalars. To this end, we notice that the decays $S,A \rightarrow \bar q_i q_j$ will be governed by the elements of $N_u$ and $N_d$ which are now almost fixed because $V_R$ is approximately defined in \Eqn{RHckm}. This is a consequence of the reducible Yukawa parameters structure of the x2HDMs, leaving all flavor couplings to be governed by $V_L$ and $V_R$. Thus, we can wonder what are the effects of flavor data in the non-SM scalar branching ratios. By taking \Eqn{RHckm}, we are fully equipped to compute the relevant two-body scalar decays into a quark anti-quark pair. For benchmark values of $M_S=M_A=1.5$ TeV, the results are shown in Table~\ref{tab:decays} for x2HDM-2, where the FCNCs are independent of $\tan\beta$, leading to fixed values of the branching ratios for any particular $V_R$. The results for x2HDM-1 are shown in Figure~\ref{fig:decays}, due to the explicit dependence on $\tan\beta$. In the case of x2HDM-2, the nonstandard scalars will preferably decay into down-type quarks, because the couplings are proportional to the up-type masses, whereas the up-type decays are proportional to the down-type masses, as seen in \Eqn{e:NuNdx2}. For the x2HDM-1, the same does not necessarily hold, as there are two contributions for flavor-diagonal decays, as shown in \Eqn{e:NuNdx1}. The different dependence on $\tan\beta$ of both contributions will make the $S \rightarrow \bar t t$ or $S \rightarrow \bar b b$ predominance be fully determined by $\tan\beta$. In fact, in the x2HDM-1 we find that the $\bar b b$ final state only surpasses $\bar t t$ for values of $\tan\beta \gtrsim 10$. In the limit $\tan\beta\to\infty$, the Yukawa couplings of x2HDM-1 and x2HDM-2 are equal, apart from sign differences which are irrelevant here, as can be seen from Eqs.~\eqref{e:NuNdx1} and \eqref{e:NuNdx2}. We can see the predominance of $\bar b b$ decays over $\bar t t$ in Table~\ref{tab:decays}. From Figure~\ref{fig:decays}, we see that the $\bar b s$ final state dominates over the $\bar b b$. This can easily be understood in the light of $N_d$, where for the particular example of $V_R$ used, the entry $\left(N_d\right)_{32}$ is one order of magnitude above $\left(N_d\right)_{33}$. This occurs because of the large $(23)$ element of $V_R$, as shown in \Eqn{RHckm}, which means a large $ts$ element of $V_R$, because of which the $s$ quark channel gets an enhancement with respect to the $b$ quark channels from the $t$-quark mass. By inspecting Figure~\ref{fig:decays}, we can also see the interplay between the two terms of \Eqn{e:NuNdx1}, which gives different behaviors to $S\rightarrow \bar{t} t$ and $S\rightarrow \bar{t} c$.
%
\iffalse
\begin{table}[h]
\centering
\begin{tabular} {||c||c|c|c|c|c||}
\hline  \\ [-1.1em] 
$\multirow{2}{*}{\text{x2HDM-2}}$ &
 $\displaystyle{\frac{S \rightarrow \text{ss}}{S \rightarrow \text{bb}}}$ & $ \displaystyle{\frac{S \rightarrow  \text{bs}}{S\rightarrow \text{bb}}} $&$ \displaystyle{\frac{S \rightarrow \text{cc}}{S \rightarrow  \text{bb}}}$ & $ \displaystyle{\frac{S \rightarrow \text{tc}}{S \rightarrow  \text{bb}}}$  & $ \displaystyle{\frac{S \rightarrow \text{tt}}{S \rightarrow  \text{bb}}}$ \\

\\[-1em]
 
\cline{2-6} 
\\[-1em]
&$ 6\times10^{-2} $&$ 17 $&$ 4\times10^{-5} $ & $10^{-2}$ & $5\times10^{-4}$ \\
 \hline
\end{tabular}
\caption{\label{tab:decays} \small Relative branching ratios for the two body fermionic decays of $S$ and $A$ for the x2HDM-2, normalized by the branching ratio of the decay into a $\bar b b$ pair, for $M_S=M_A=1.5 \text{ TeV}$. One of the quarks in each process is to be taken to be an antiquark. We have not marked which one, because the result is independent of this choice. We have checked that the relative branching ratios for $A$ are of the same order of magnitude.}
\end{table}
\fi
%

%
\begin{table}[h]
\centering
\begin{tabular} {||c||c|c|c|c|c||}
\hline  \\ [-1.1em] 
$\text{x2HDM-2}$ &
 $\displaystyle{\frac{H \rightarrow \text{ss}}{H \rightarrow \text{bb}}}$ & $ \displaystyle{\frac{H \rightarrow  \text{bs}}{H\rightarrow \text{bb}}} $&$ \displaystyle{\frac{H \rightarrow \text{cc}}{H \rightarrow  \text{bb}}}$ & $ \displaystyle{\frac{H \rightarrow \text{tc}}{H \rightarrow  \text{bb}}}$  & $ \displaystyle{\frac{H \rightarrow \text{tt}}{H \rightarrow  \text{bb}}}$ \\

\\[-1em]
 
\cline{1-6} 
\\[-1em]
$H\equiv S$ &\multirow{2}{*}{$ 6\times10^{-2} $} &\multirow{2}{*}{$ 17 $} &\multirow{2}{*}{$ 4\times10^{-5} $} & \multirow{2}{*}{$10^{-2}$} & $5\times10^{-4}$ \\
%\cline{1-1}
%\\[-1.1em]

$H\equiv A$ &&& & & $6\times10^{-4}$ \\
 \hline
\end{tabular}
\caption{\label{tab:decays} \small Relative branching ratios for the two body fermionic decays of $S$ and $A$ for the x2HDM-2, normalized by the branching ratio of the decay into a $\bar b b$ pair, for $M_S=M_A=1.5 \text{ TeV}$. One of the quarks in each process is to be taken to be an antiquark. We have not marked which one, because the result is independent of this choice. }
\end{table}

The features shown in this section are distinctive characteristic of the x2HDMs, which can be used to falsify the models here shown.

\section{Summary}
\label{s:summary}
In this paper, we studied some properties of a minimal extension of
the SM, the 2HDM, and, in particular, focused on the study of the
FCNCs of the model.  We studied different variants of the 2HDM,
resulting from the imposition of different symmetries on the Yukawa
interactions of the model.  Since these symmetries do not commute with
the full gauge group of the SM Lagrangian (in practice, they are
broken by the hypercharge), the imposed symmetries are effectively
approximate symmetries of the theory.  Taking advantage that the
renormalization group equations of the 2HDM are well-known, assuming
the symmetry to be a true symmetry of the Lagrangian at an energy
scale $\mu$, it is possible to compute the evolution of the deviation
from the symmetric situation with respect to the energy scale.

We advocate the bidoublet notation, widely popular in the context of
left-right symmetric theories, applied to the context of the
2HDM.  While it may seem a convoluted exercise which increases the
problem's complexity, we argue for its benefits.  Namely, imposing
simple symmetries on the bidoublet, we are able to recover the
paradigmatic type-I and type-II 2HDM models, as well as formulate two
new 2HDM variants, which until now remained unstudied.

Through this paper, our main goal was to search for models where the
general arbitrariness of the FCNC couplings was reduced, following the
motivation of BGL models, by relating these couplings with the quark
mixing matrices.  We find a class of new models, where the FCNCs are
controlled by the left- and right-handed quark mixings.  Due to the
particular relations between the Yukawa matrices of the model, we name
this new class of models the crossed 2HDM (x2HDM).  In one of these
such models, the x2HDM-1, we show that it is possible to impose a
symmetry on the Yukawa sector such that the FCNCs are fully controlled
by the left- and right-handed CKMs, as well as the ratio between the
scalar doublets VEVs.  We also point out that, while this symmetry is
approximate in the 2HDM context, it is automatically imposed when
dealing with the LRSM.  As such, this model can be taken as the
electroweak scale incarnation of the LRSM, given that the LRSM relies
on a 2HDM structure.  Following up on this intimate connection between
the x2HDM-1 and the LRSM, a comprehensive flavor analysis, when
paired up with the RGE study, may lead to valuable insight on the
validity of some LRSMs, which we wish to explore in a future
work.  Furthermore, it is important to note that some of the
conclusions obtained for the x2HDM-1 are equally valid or extendable
to the LRSM, such as the excluded region for the Higgs-doublets' (the
bidoublet's in the LRSM context) VEVs.  We also present a second model,
dubbed x2HDM-2, where the FCNC structure is further simplified, being
entirely controlled by the left- and right-handed CKMs, independent of
the VEV ratio.  While we do not present a UV-completion for this
model, we consider this model as a valuable argument for the benefits
of a change of outlook (in this case, notations), to uncover new
interesting possibilities.

We have also performed a phenomenological analysis of the x2HDMs, 
to showcase their predictive power.  In the paradigm of the 
alignment limit, as well as assuming $M_S=M_A$, the tree-level 
contributions to the $\Delta F=2$ processes are simplified, but still 
remain.  As such, the restrictive flavor data on $\Delta M_K$, 
$\Delta M_{B_s}$, and $\Delta M_{B_d}$, constrain the model.  
However, the same couplings are responsible not only for the neutral 
meson oscillations, but also for other flavor processes such as the 
two-body fermionic decays of the nonstandard scalars of the theory.  
As such, a specific example for $V_R$ is shown, which was obtained by 
requiring the compatibility of the models with $\Delta F=2$ data. It leads
to specific values for the branching ratios of both $S$ and $A$ for the 
x2HDM-2, and a distinctive pattern of these quantities as a function of $\tan\beta$ 
for the x2HDM of type 1.

As a final note, hopefully, the explicit relation between the x2HDM-1
and the LRSM, together with the economical structure of the FCNCs of
both x2HDMs, as well as the benefits of a change in notation for
uncovering models, will lead to a renewed aesthetic motivation for the
study of 2HDMs, apart from the supersymmetric embedding.

\section*{Acknowledgments}
The work of GB was supported by Funda\c{c}\~{a}o para a Ci\^{e}ncia e
a Tecnologia (FCT, Portugal) through the projects CFTP-FCT Unit 777
(UID/FIS/00777/2013 and UID/FIS/00777/2019), CERN/FIS-PAR/0004/2017,
and PTDC/FIS-PAR/29436/2017 which are partially funded through POCTI
(FEDER), COMPETE, QREN and EU.  The work of ML is funded by
Funda\c{c}\~ao para a Ci\^encia e Tecnologia-FCT Grant
No.PD/BD/150488/2019, in the framework of the Doctoral Programme
IDPASC-PT.  PBP's research was supported by the SERB grant
EMR/2017/001434 of the Government of India.  DD and PBP gratefully
acknowledge the warm hospitality of CFTP, Lisbon where part of this
work was done.

%%%%%%%%%%%%%%%%%   References %%%%%%%%%%%%%%%%%%%%%%%%%%%%%%%%%%%%

\bibliographystyle{JHEP}
\bibliography{2hdm.bib}

\providecommand{\href}[2]{#2}\begingroup\raggedright\begin{thebibliography}{10}

\bibitem{Branco:2011iw}
G.~Branco, P.~Ferreira, L.~Lavoura, M.~Rebelo, M.~Sher, and J.~P. Silva, {\it
  {Theory and phenomenology of two-Higgs-doublet models}},  {\em Phys.\ Rept.}
  {\bf 516} (2012) 1--102, [\href{http://arxiv.org/abs/1106.0034}{{\tt
  arXiv:1106.0034}}].

\bibitem{Bhattacharyya:2015nca}
G.~Bhattacharyya and D.~Das, {\it {Scalar sector of two-Higgs-doublet models: A
  minireview}},  {\em Pramana} {\bf 87} (2016), no.~3 40,
  [\href{http://arxiv.org/abs/1507.06424}{{\tt arXiv:1507.06424}}].

\bibitem{Glashow:1976nt}
S.~L. Glashow and S.~Weinberg, {\it {Natural Conservation Laws for Neutral
  Currents}},  {\em Phys. Rev.} {\bf D15} (1977) 1958.

\bibitem{Das:2018qjb}
D.~Das, {\it {2HDM without FCNC: off the beaten tracks}},  {\em Eur.\ Phys.\
  J.\ C} {\bf 78} (2018), no.~8 650,
  [\href{http://arxiv.org/abs/1803.09430}{{\tt arXiv:1803.09430}}].

\bibitem{Botella:2018gzy}
F.~J. Botella, F.~Cornet-Gomez, and M.~Nebot, {\it {Flavor conservation in
  two-Higgs-doublet models}},  {\em Phys. Rev. D} {\bf 98} (2018), no.~3
  035046, [\href{http://arxiv.org/abs/1803.08521}{{\tt arXiv:1803.08521}}].

\bibitem{Branco:1996bq}
G.~C. Branco, W.~Grimus, and L.~Lavoura, {\it {Relating the scalar flavor
  changing neutral couplings to the CKM matrix}},  {\em Phys. Lett.} {\bf B380}
  (1996) 119--126, [\href{http://arxiv.org/abs/hep-ph/9601383}{{\tt
  hep-ph/9601383}}].

\bibitem{Bhattacharyya:2014nja}
G.~Bhattacharyya, D.~Das, and A.~Kundu, {\it {Feasibility of light scalars in a
  class of two-Higgs-doublet models and their decay signatures}},  {\em Phys.\
  Rev.\ D} {\bf 89} (2014) 095029, [\href{http://arxiv.org/abs/1402.0364}{{\tt
  arXiv:1402.0364}}].

\bibitem{Botella:2014ska}
F.~Botella, G.~Branco, A.~Carmona, M.~Nebot, L.~Pedro, and M.~Rebelo, {\it
  {Physical Constraints on a Class of Two-Higgs Doublet Models with FCNC at
  tree level}},  {\em JHEP} {\bf 07} (2014) 078,
  [\href{http://arxiv.org/abs/1401.6147}{{\tt arXiv:1401.6147}}].

\bibitem{Mohapatra:2019qid}
R.~N. Mohapatra, G.~Yan, and Y.~Zhang, {\it {Ameliorating Higgs induced flavor
  constraints on TeV scale $W_R$}},  {\em Nucl. Phys. B} {\bf 948} (2019)
  114764, [\href{http://arxiv.org/abs/1902.08601}{{\tt arXiv:1902.08601}}].

\bibitem{Bhattacharyya:2013rya}
G.~Bhattacharyya, D.~Das, P.~B. Pal, and M.~Rebelo, {\it {Scalar sector
  properties of two-Higgs-doublet models with a global U(1) symmetry}},  {\em
  JHEP} {\bf 10} (2013) 081, [\href{http://arxiv.org/abs/1308.4297}{{\tt
  arXiv:1308.4297}}].

\bibitem{Das:2015mwa}
D.~Das and I.~Saha, {\it {Search for a stable alignment limit in
  two-Higgs-doublet models}},  {\em Phys.\ Rev.\ D} {\bf 91} (2015), no.~9
  095024, [\href{http://arxiv.org/abs/1503.02135}{{\tt arXiv:1503.02135}}].

\bibitem{Dev:2014yca}
P.~Bhupal~Dev and A.~Pilaftsis, {\it {Maximally Symmetric Two Higgs Doublet
  Model with Natural Standard Model Alignment}},  {\em JHEP} {\bf 12} (2014)
  024, [\href{http://arxiv.org/abs/1408.3405}{{\tt arXiv:1408.3405}}].
  [Erratum: JHEP 11, 147 (2015)].

\bibitem{Das:2019yad}
D.~Das and I.~Saha, {\it {Alignment limit in three Higgs-doublet models}},
  {\em Phys.\ Rev.\ D} {\bf 100} (2019), no.~3 035021,
  [\href{http://arxiv.org/abs/1904.03970}{{\tt arXiv:1904.03970}}].

\bibitem{Botella:2009pq}
F.~J. Botella, G.~C. Branco, and M.~N. Rebelo, {\it {Minimal Flavour Violation
  and Multi-Higgs Models}},  {\em Phys. Lett.} {\bf B687} (2010) 194--200,
  [\href{http://arxiv.org/abs/0911.1753}{{\tt arXiv:0911.1753}}].

\bibitem{Mohapatra:1974gc}
R.~Mohapatra and J.~C. Pati, {\it {A Natural Left-Right Symmetry}},  {\em
  Phys.\ Rev.\ D} {\bf 11} (1975) 2558.

\bibitem{Mohapatra:1974hk}
R.~N. Mohapatra and J.~C. Pati, {\it {Left-Right Gauge Symmetry and an
  Isoconjugate Model of CP Violation}},  {\em Phys.\ Rev.\ D} {\bf 11} (1975)
  566--571.

\bibitem{Senjanovic:1975rk}
G.~Senjanovi\'c and R.~N. Mohapatra, {\it {Exact Left-Right Symmetry and
  Spontaneous Violation of Parity}},  {\em Phys.\ Rev.\ D} {\bf 12} (1975)
  1502.

\bibitem{Deshpande:1990ip}
N.~G. Deshpande, J.~F. Gunion, B.~Kayser, and F.~I. Olness, {\it {Left-right
  symmetric electroweak models with triplet Higgs}},  {\em Phys. Rev.} {\bf
  D44} (1991) 837--858.

\bibitem{Atwood:1996vj}
D.~Atwood, L.~Reina, and A.~Soni, {\it {Phenomenology of two Higgs doublet
  models with flavor changing neutral currents}},  {\em Phys. Rev. D} {\bf 55}
  (1997) 3156--3176, [\href{http://arxiv.org/abs/hep-ph/9609279}{{\tt
  hep-ph/9609279}}].

\bibitem{Nebot:2015wsa}
M.~Nebot and J.~P. Silva, {\it {Self-cancellation of a scalar in neutral meson
  mixing and implications for the LHC}},  {\em Phys. Rev. D} {\bf 92} (2015),
  no.~8 085010, [\href{http://arxiv.org/abs/1507.07941}{{\tt
  arXiv:1507.07941}}].

\bibitem{Wolfenstein:1983yz}
L.~Wolfenstein, {\it {Parametrization of the Kobayashi-Maskawa Matrix}},  {\em
  Phys. Rev. Lett.} {\bf 51} (1983) 1945.

\end{thebibliography}\endgroup

\end{document}